\RequirePackage{fix-cm}
\documentclass[smallextended]{svjour3}       
\smartqed  
\usepackage{appendix}
\usepackage{graphicx}
\usepackage{amssymb}
\usepackage{color}
\newcommand{\etal}{  {\it et al.}}

\newcommand{\mnras}{  {\it Mon. Not. Roy. Astron. Soc.}}

\newcommand{\solphys}{{\it Solar Phys.}}

\newcommand{\astn}{   {\it Astron. Nachr.}} 

\begin{document}

\title{A digital method to calculate the true areas \\of sunspot groups
}

\author{H. \c{C}akmak}

\titlerunning{A digital method to calculate the true areas of sunspot groups}

\authorrunning{H. \c{C}akmak} 

\institute{H. \c{C}akmak\at
  Istanbul University Science Faculty, Astronomy and Space Science Department \\
  34119 Beyazit / Istanbul - Turkey\\
  \email{hcakmak@istanbul.edu.tr}           
}

\date{Received: date / Accepted: date}

\maketitle

\begin{abstract}
The areas of sunspots are the most prominent feature of the development of sunspot groups. Since 
the size of sunspot areas depend on the strength of the magnetic field, accurate measurements of 
these areas are important. In this study, a method which allows to measure true areas of the sunspots 
is introduced. A Stonyhurst disk is created by using a computer program and is coincided with solar 
images. By doing this, an accurate heliographic coordinate system is formed. Then, the true area of 
the whole sunspot group is calculated in square degrees with the aid of the heliographic coordinates 
of each picture element forming the image of the sunspot group.

This technique's use is not limited with sunspot areas only. The areas of the flare and filaments 
observed on the chromospheric disk can also be calculated with the same method. In addition to 
this, it is possible to calculate the area of any occurrence on the solar disk, whether it is 
related to an activity or not.
\end{abstract}

\section{Introduction}
     \label{S-Introduction} 
Sunspots are the most obvious feature of solar magnetic activity. To understand the 
development of the solar activity, revealing the morphologic and kinematic behaviors of the 
sunspots on the solar surface is required. Therefore, analyzing the emergence patterns, 
developments and decay of the sunspots on the solar surface are the most important steps to 
constitute the sunspot group models. Evolution of the groups on the surface is observed with 
the evolution the both umbral and penumbral areas \cite{Gafeira2012}, \cite{Hathaway2008}. 
McIntosh \cite{McIntosh1990} described the classification of the sunspot groups depending on the 
appearance and the area covered on the surface. Here, the areas of the sunspots are an 
important criterion and they enable the groups to be distinguished from each other.

The size and distribution of the sunspots are an indication of the complexity of the activity 
field which the group is in \cite{Zirin1988}. Hence, the positions of the sunspots in 
the group also show the magnetic field distribution. Because the size of both umbral and 
penumbral areas of the sunspots are proportional to the magnitude of the magnetic field 
strength \cite{Schlichenmaier2010}, the sunspot areas which are reaching a certain size or 
being disintegrated or tending to merge with other sunspots are indicating the different 
phases of the group development \cite{McIntosh1990}. An accurate measurement of the sunspot 
areas, therefore, could provide important information.

Areas and heliographic positions of the sunspot groups were regularly calculated and archived 
at the Royal Greenwich Observatory from 1874 until 1976. The results were published in Greenwich 
Photoheliographic Results. After 1976, Debrecen Heliophysical Observatory (DHO) took over 
this mission. Now, the daily data about areas and positions of the sunspot groups are published 
in Debrecen Photoheliographic Data. Video images of the sunspot groups are used for area 
measurement and an isodensity line is fitted to the edge of the spot at DHO. The sunspot 
group areas are calculated as follows: the area is divided into small squares with a grid 
system, then, the number of the squares in the area are counted and added up. And then, the 
total area is transformed into area on the solar disk \cite{Gyori1998}, \cite{Sarychev2006}.

Many researchers are using the circular marking method in which the area of a sunspot is 
determined with the area of a circle superimposed on it. All of the sunspot areas in the 
group are individually calculated and their sum will give the total area of the sunspot group 
\cite{Meadows2002}, \cite{Arlt2013}. The area of the sunspot group $\it{A_M}$, on the solar disk is 
calculated by
   \begin{eqnarray}
	A_M &=& \frac{2 A_S 10^6}{\pi D^2 \cos(\rho)} \nonumber
   \end{eqnarray}
\noindent
where $\it{A_S}$ is the measured size of a sunspot group in the image, D is the diameter of the 
image, $\rho$ is the angular distance of the sunspot group's center from the center of 
the disk. The area of the sunspot group is given in millionths of the apparent solar disk in 
these studies and includes the correction for foreshortening. Whereas in the method explained 
in here, the area of the group is calculated in square degrees.

Nowadays, automated sunspot recognition techniques are developed and the researches are almost 
concentrated in this field instead of the development the old studies which have hand-made 
measurements. A Software package called SAM at DHO is used for automated recognition. At 
the University of Bredford, an automated program is used to produce Solar Feature Catalogue. 
Another automated program called StarTool is applied to digital images \cite{Gyori2005}. 
The articles written by Gy\H{o}ri \cite{Gyori1998} and Fonte \& Fernandes \cite{Fonte2009} 
have detailed descriptions about automated recognition to determine the edge of the sunspots 
with the image processing techniques. But this approach brings some mistakes on the boundaries 
of the sunspots due to the blurring and smoothing processes and wipes out some parts of the group, 
especially small sunspots close to big penumbral structures and some umbral spots close to each 
other. Therefore, it can be said that these techniques are rough estimates and can not give 
true areas of the sunspot groups. Semi-automated approaches may solve these points by 
adjusting the threshold values visually.

\section{The method} 
      \label{S-method}      
The first step to measure the sunspot areas in digital environment is to mark the sunspot 
areas on the solar disk image. Initially, disk images are transformed to the RGB grayscale 
format which has an intensity range from 0 to 255. Intensity value 0 means black, 255 means white 
color in RGB format. This increases the accuracy of the edge detection in the next step. 
The edges of the sunspots are determined by using the {\it Contour Trace} algorithm. Ren\etal 
\cite{Ren2002} and Wagenknecht \cite{Wagenknecht2007} gave detailed information about this 
technicque. In this procedure, isodensity lines are drawn around sunspots by changing the threshold 
intensity level. Pixels which have intensity value higher than threshold value are taken into account 
and processed for contouring. 

Two different intensity levels are shown in Fig. 1 as an example. In the figure, the sunspots have 
lines on their edges, but some areas which are not sunspots have border as well. This is the hard 
point to decide and carry out. When higher threshold values are selected, more bordered areas are 
seen and the areas of the sunspots are wider. After a proper value is selected, the necessary 
corrections have to be made by removing the areas which are not sunspots and holes in the sunspots. 
When these are done, sunspot group will be seen as in Fig. 2b. Afterwards, the image is made ready 
for area calculation by filling the areas of the sunspots with black in the contour drawings (Fig. 2c).

Any image in the digital media consists of many picture elements, called pixels. With this approach, 
the size of the pixel area can be calculated easily in square degrees, when the heliographic coordinates 
of every pixel are known in the solar disk image. To achieve this, it will be enough to superimpose a 
\begin{figure}[h]    
   \centerline{
	\includegraphics[width=0.495\textwidth,clip=]{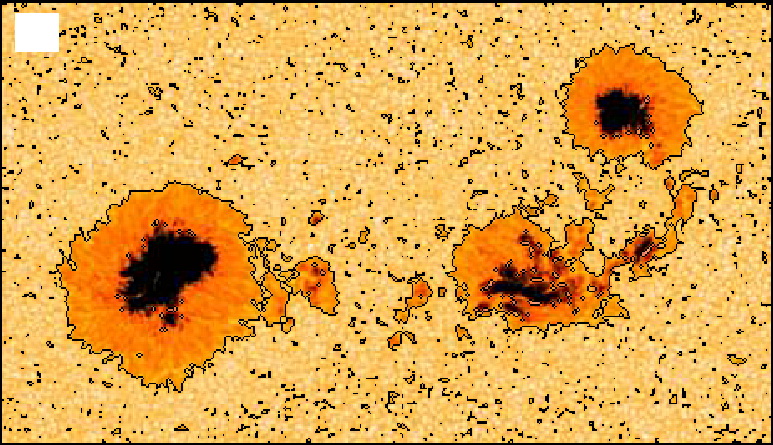}
        \includegraphics[width=0.495\textwidth,clip=]{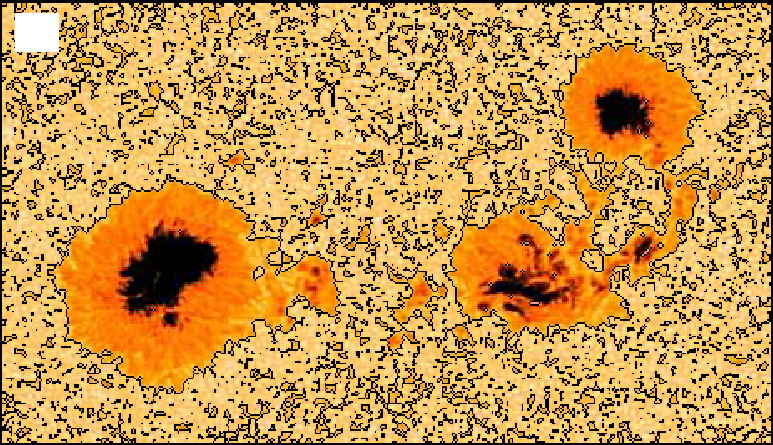}
   }
   \vspace{-0.285\textwidth}
   \hspace{ 0.008\textwidth}  \color{black} \small {a}
   \hspace{0.475\textwidth}    \color{black} \small {b}
   \vspace{0.245\textwidth}
   \caption{Isodensity lines of a sunspot group with different threshold intensity values. 
	    Intensity levels are (a) 155 and (b) 165, respectively.}
   \label{F-One}
\end{figure}
\begin{figure}[h]    
   \centerline{
	\includegraphics[width=0.325\textwidth,clip=]{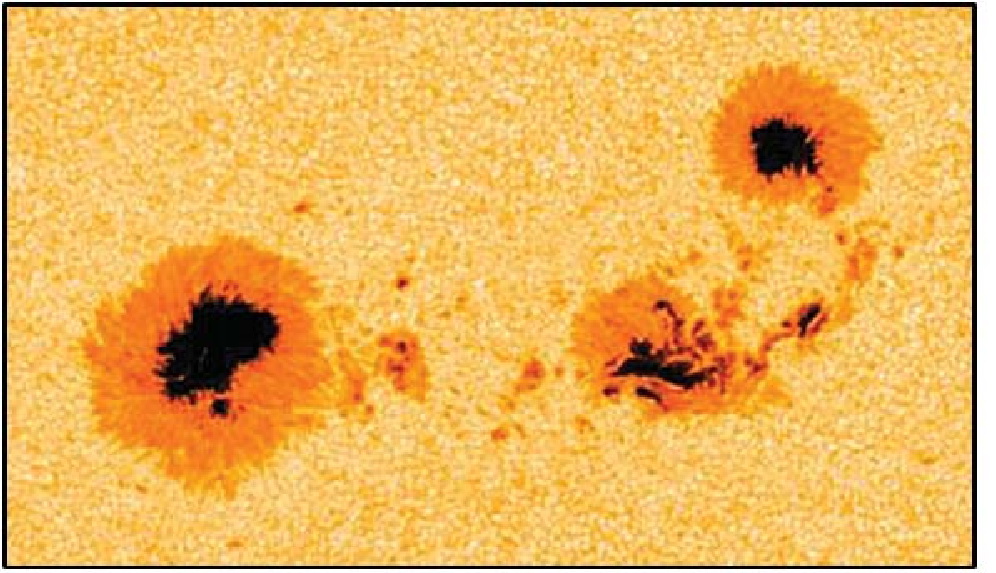}
        \includegraphics[width=0.325\textwidth,clip=]{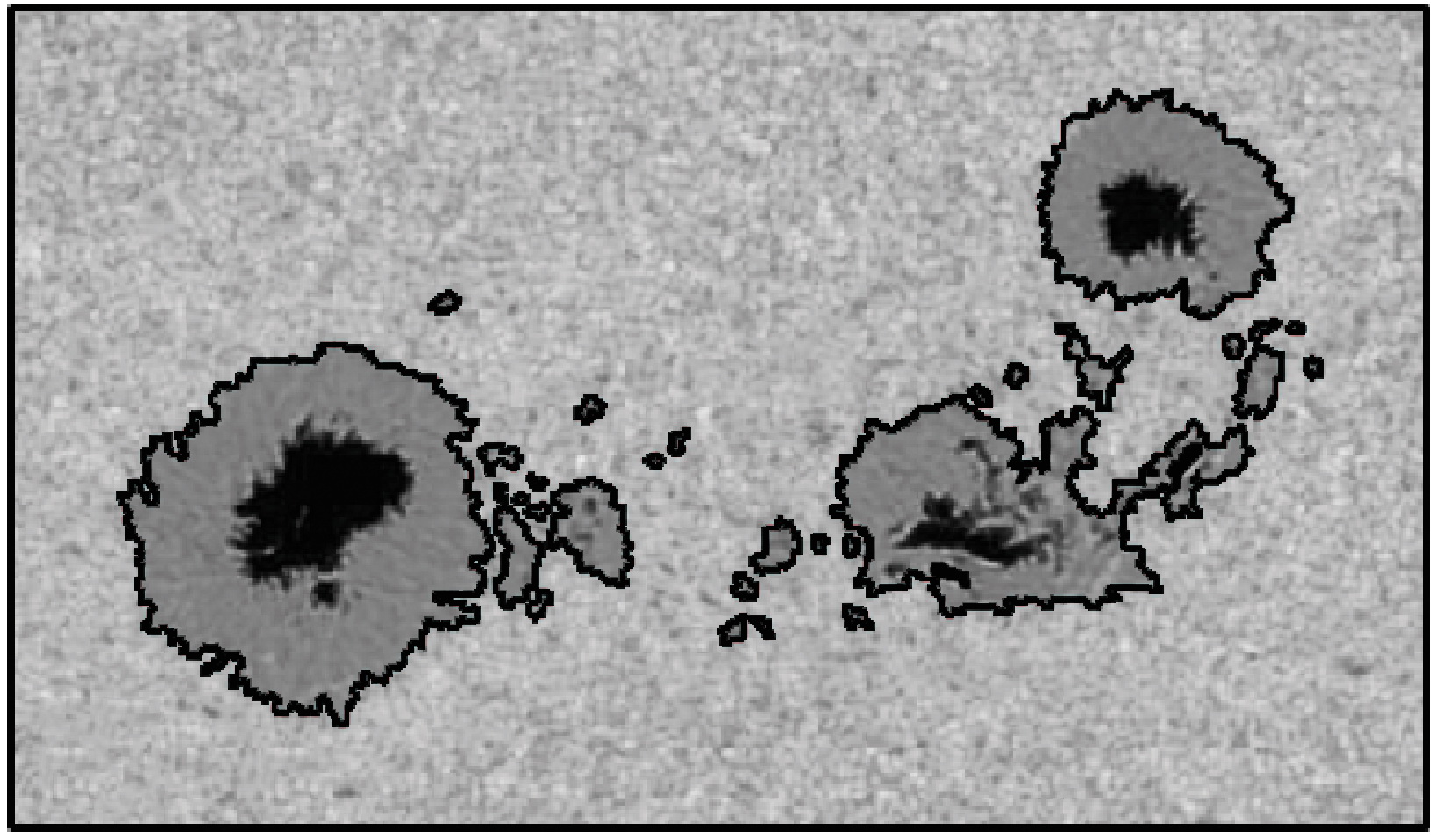}
        \includegraphics[width=0.325\textwidth,clip=]{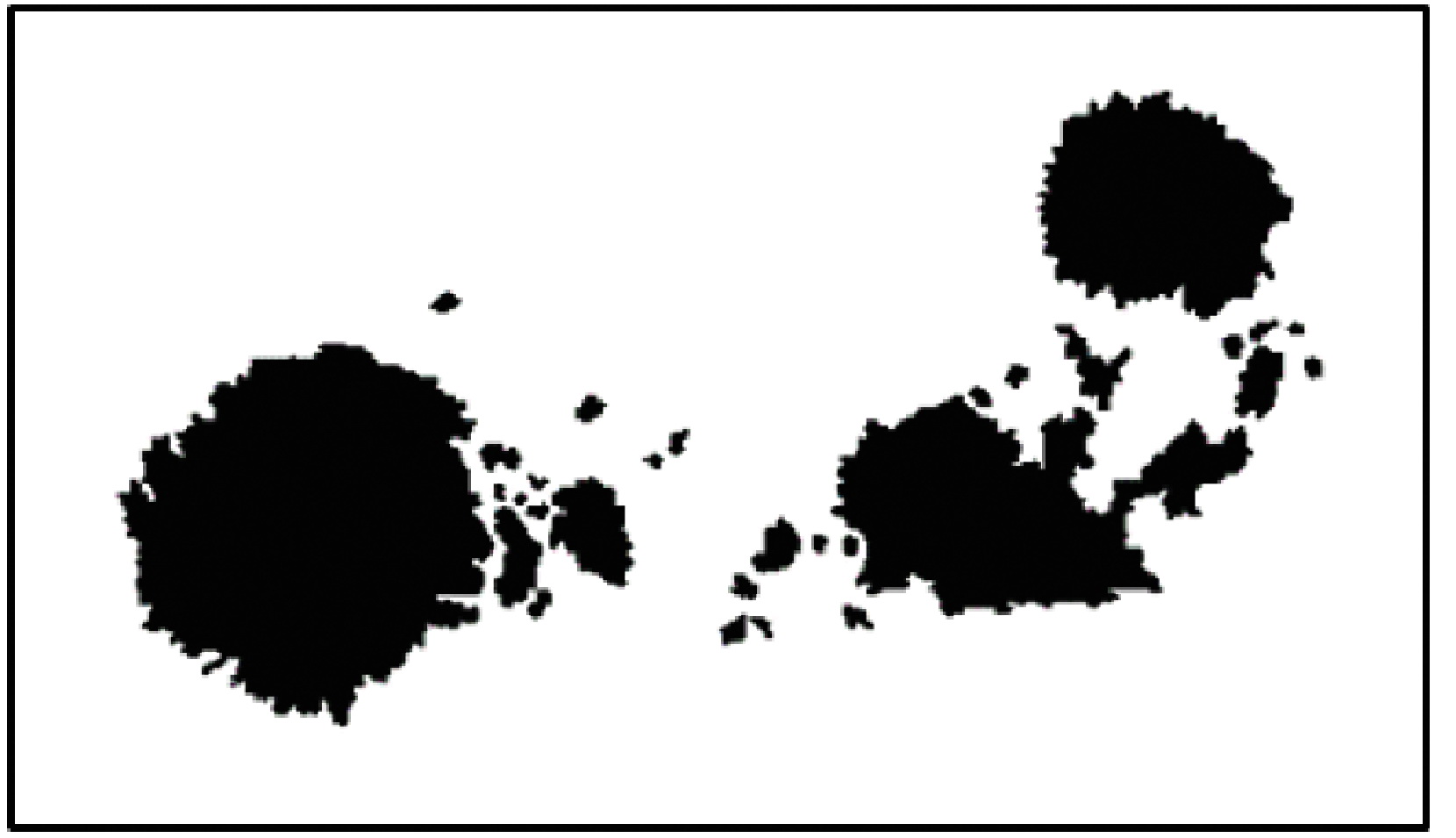}
   }
   
   \vspace{-0.045\textwidth}
   \hspace{ 0.004\textwidth}  \color{black} \small {(a)}
   \hspace{0.283\textwidth}    \color{black} \small {(b)}
   \hspace{0.285\textwidth}    \color{black} \small {(c)}

   \caption{(a) A solar sunspot group image taken by Solar Dynamics Observatory\protect\footnotemark[1], 
	    (b) group image in black and white; black line shows detected edges, 
            (c) group image with filled areas.}
   \label{F-Two}
\end{figure}
Stonyhurst disk (for detail see Cortie \cite{Cortie1908}) on the disk image. Solar parameters (position 
angle $\it{P}$, equatorial angle $\it{B_0}$ and initial longitude angle $\it{L_0}$) at the time of the 
observation need to be calculated by using the astronomical almanac to prepare this Stonyhurst disk 
(hereafter referred as graticule). 
\footnotetext[1]{http://sdo.gsfc.nasa.gov/.}

Using these parameters, the graticule can be placed over the desired image with the aid of a computer 
software (for detail see \cite{Cakmak2010}). As an example, a HMIIF (Helioseismic and Magnetic Imager 
Intensitygram - Flattened) solar disk image is processed in accordance with the steps described above 
(Fig. 3a) and the graticule is placed on the disk using software (Fig. 3b). The image was taken by the 
Solar Dynamics Observatory (SDO) satellite in 15.06.2012.
\begin{figure}[h] 
   \centerline{	\includegraphics[width=1.0\textwidth,clip=]{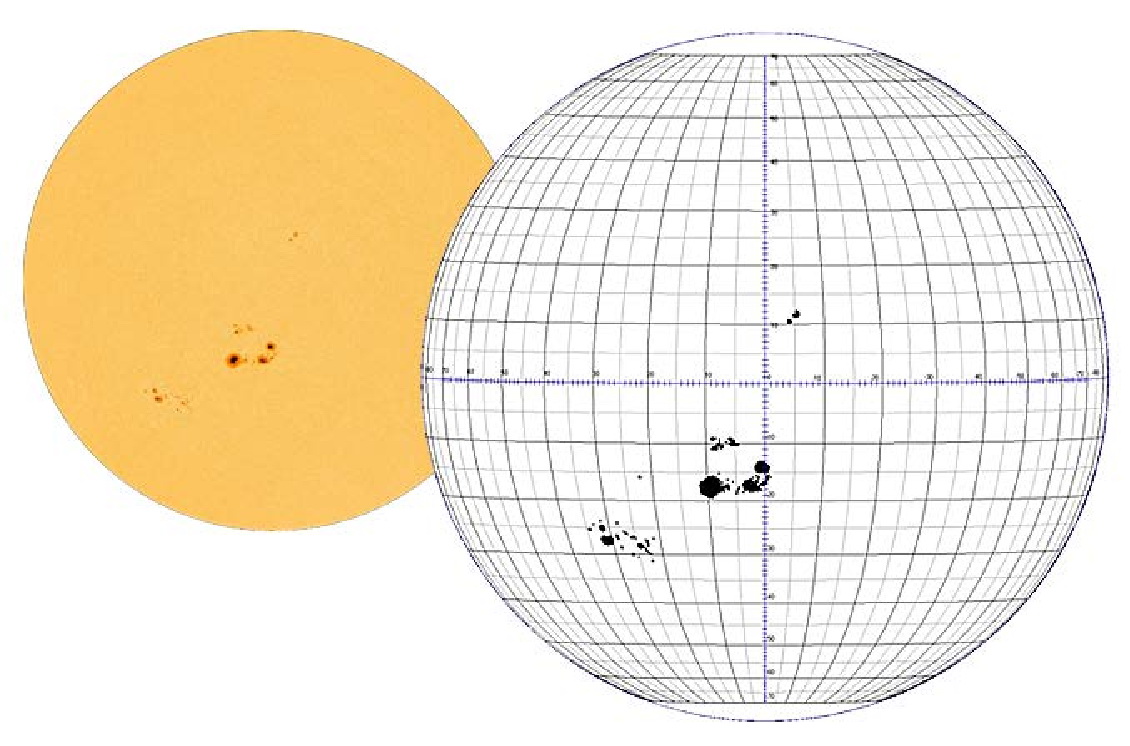} }

   \vspace{-0.637\textwidth}
   \hspace{0.225\textwidth} \color{black} \small {N}\\

   \vspace{0.15\textwidth}
   \hspace{0.015\textwidth} \color{black} \small {E}\\
   
   \vspace{0.1\textwidth}
   \hspace{0.03\textwidth} \color{black} \small {(a)}\\
   
   \vspace{0.125\textwidth}
   \hspace{0.45 \textwidth}   \color{black} \small {(b)}
   
   \vspace{0.028\textwidth}

   \caption{(a) A SDO HMIIF solar image in 15.06.2012. Courtesy of NASA/SDO and the AIA, EVE, and 
               HMI science teams. (b) Graticule system placed on the disk with filled sunspot areas.}
   \label{F-three}
\end{figure}

\subsection{Calculation of the heliographic coordinates of a pixel} 
  \label{S-pixelcoordinates}

Three-dimensional coordinates system for a sphere were used in the preparation of the graticule system in the computer environment (Fig. 4). Here, the cartesian coordinates of a point using it's spherical coordinates are given by
\begin{eqnarray} \label{Eq-cartesian}
	x &=& r \cos B \sin L , \\
	y &=& r \cos B \cos L , \nonumber \\
	z &=& r \sin B , 
\end{eqnarray}
where {\it r} is the radius of the sphere (solar disk image), {\it B} is the latitude angle and {\it L} is 
the longitude angle \cite{SmartG77}. Also, in order to calculate the apparent heliographic coordinates 
of a point on the graticule , the angle {\it B} is selected between +90$^\circ$ (for Northern emisphere) and 
-90$^\circ$ (for Southern hemisphere) and angle {\it L} is selected between +90$^\circ$ (for West hemisphere) 
and -90$^\circ$ (for East hemisphere). Then, these three-dimensional coordinates have to be converted to 
two-dimensional coordinates to draw on the image (depending on {\it P} and {\it $B_o$} parameters of the Sun). 
These are done with the well known transformation and projection equations \cite{Stephens00}, \cite{GovilPai04}, 
\cite{Wright00}. As seen in Fig. 4, by rotating {\it x}-axis {\it $B_o$} degrees and {\it y}-axis {\it P} 
degrees, the projection coordinates {\it $x_p$} and {\it $y_p$} of a point are determined by
\begin{eqnarray} \label{Eq-projection}
	x_p &=& x_o + (z \cos B_o - y \sin B_o) \sin P + x \cos P , \\
	y_p &=& y_o - (z \cos B_o - y \sin B_o) \cos P - x \sin P , 
\end{eqnarray}
where $x_o$ and $y_o$ are the coordinates of the disk image center.

\begin{figure}[b]     
   \centerline{
     \includegraphics[width=0.4\textwidth,clip=]{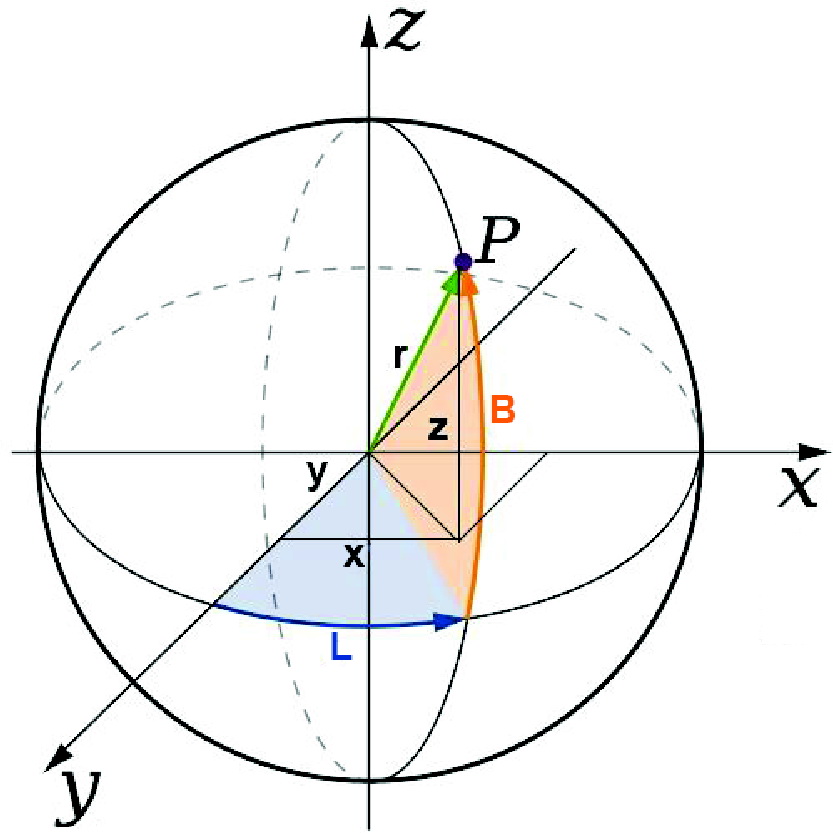}
   }
   \caption{Three-dimensional cartesian ($\it {x, y, z}$) and spherical ($\it r$ radius, $\it B$ latitude 
and $\it L$ longitude angle) coordinates of a point.}
   \label{F-four}
\end{figure}

From this point, the heliographic coordinates of a pixel can be calculated by the inverse process. 
All coordinates are being given in pixels in the following process. As defined above, {\it $x_p$} and 
{\it $y_p$} are the screen coordinates of a pixel taken into account and let {\it $x_a$} and {\it $y_a$} 
be the Cartesian coordinates of that pixel with relative to the solar disk image center. From the 
equations (3) and (4), we define {\it $x_a$} and {\it $y_a$} as follows
\begin{eqnarray} \label{Eq-screen}
	x_a &=& x_p - x_o , \nonumber \\
	y_a &=& y_o - y_p , \nonumber \\
	x_a &=& z_r \sin P + x \cos P , \\
	y_a &=& z_r \cos P - x \sin P , 
\end{eqnarray}
where {\it $z_r$} is the transformed values of {\it $z$} and {$z_r = z \cos B_o - y \sin B_o$}. From the 
equations (5) and (6), we find the below equations by the inverse transformation.
\begin{eqnarray} \label{Eq-reverse}
	  x &=& x_a \cos P - y_a \sin P \nonumber , \\
	z_r &=& x_a \sin P + y_a \cos P \nonumber ,
\end{eqnarray}
On the other hand, we have the {\it $r^2 = {x_r}^2 + {y_r}^2 + {z_r}^2$} equation in Cartesian coordinates, 
where {\it $x_r$} and {\it $y_r$} are the transformed values of {\it $x$}, and {\it $y$}. In these transformation process {\it $x = {x_r}$}. Then,
\begin{eqnarray} \label{Eq-reverse2}
	y_r &=& (r^2 - x^2 - {z_r}^2)^{1/2} \nonumber
\end{eqnarray}
can be easily obtained. The inverse transformation for {\it $z$} is given by  
\begin{eqnarray} \label{Eq-reverse3}
	  z &=& y_r \sin B_0 + z_r \cos B_0
\end{eqnarray}
Substituting {\it $z_r$} and {\it $y_r$} into (7), we get the {\it $z$} value of the pixel in Cartesian 
coordinates. Finally, by using the equations (1) and (2), the heliographic coordinates of the pixel, 
{\it $L_p$} and {\it $B_p$}, are obtained in degrees by
\begin{eqnarray} \label{Eq-helio}
	B_p &=& \arcsin \left(\frac{z}{r}\right) \frac{180}{\pi} , \nonumber \\
	L_p &=& L_0 + \arcsin \left(\frac{x}{r \cos (90 - B_p)}\right) \frac{180}{\pi} ,\nonumber 
\end{eqnarray}
where {\it $L_0$} is the initial longitude in the heliographic coordinates at the time of the capture of the 
solar image.

\subsection{Principle of the sunspot area calculation} 
  \label{S-calculation}

Once a graticule is superimposed on a solar disk image as shown in Fig. 3b, every pixel forming the sunspot 
group (Fig. 5b) will have a heliographic latitude ($\it{B}$) and longitude ($\it{L}$) on the solar image. 
Therefore, the area of any pixel can be calculated with the help of the coordinates of its neighboring pixels. 
Let pixel-i be a pixel taken into account and let the pixel-a and pixel-b be the nearest two pixels as shown in 
Fig. 5c. The area of pixel-i can be calculated using $\Delta B$ and $\Delta L$ which are the latitudinal and 
longitudinal differences in degrees, respectively. Here, pixel-i has the same latitude with pixel-a and the 
same longitude with pixel-b. The surrounding pixels of the selected pixel are taking into account to find 
these nearest.

A computer program is used for the calculation processes. To decrease the duration of the calculations and 
speed up all processes, only a small part of the disk image is selected that the sunspot group is in 
(Fig. 5a, dotted rectangle area). Then, the heliographic coordinates of all the pixels inside the dotted 
area are calculated. The process is started with the leftmost pixel of the top row of the dotted area 
and it is continued to the right end of the row. Then, other rows from second row to the last row of the 
dotted area are processed.
\begin{figure}[b] 
     
   \centerline{
	\includegraphics[width=0.75\textwidth,clip=]{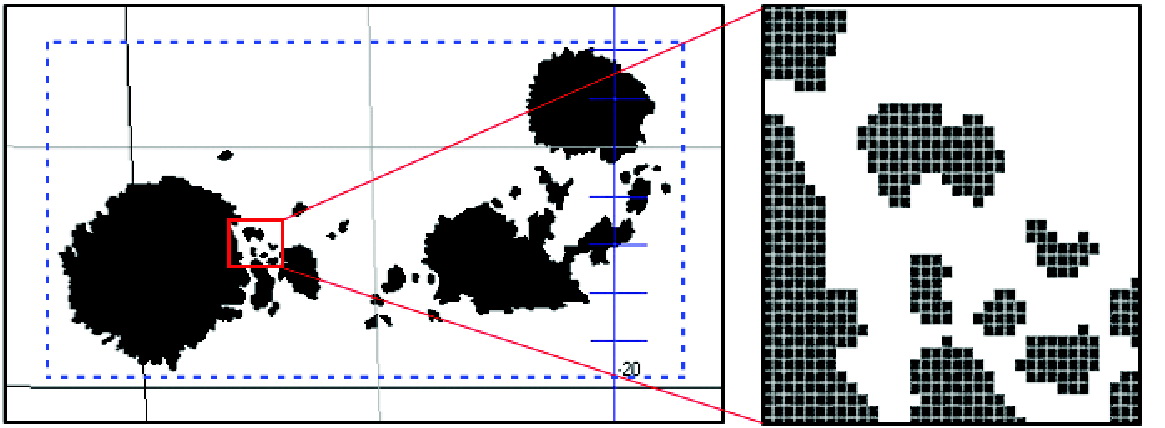}
        \includegraphics[width=0.245\textwidth,clip=]{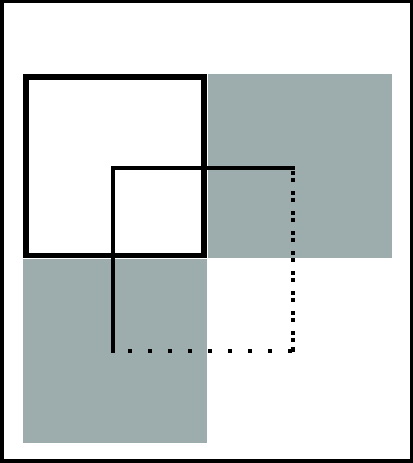}
   }
   \vspace{-0.276\textwidth}
   \hspace{-0.003 \textwidth} \color{black} \small {(a)}
   \hspace{0.635 \textwidth}   \color{black} \small {(b)}
   \hspace{0.22 \textwidth}   \color{black} \small {(c)}

   \vspace{0.04\textwidth}
   \hspace{0.87\textwidth}   \color{black} \small {$\Delta L$}
   
   \vspace{-0.02\textwidth}
   \hspace{0.79 \textwidth}   \color{black} \normalsize {$\it{i}$}
   \hspace{0.11 \textwidth}   \color{black} \normalsize {$\it{a}$}
   
   \vspace{0.05\textwidth}
   \hspace{0.77\textwidth}   \color{black} \small {$\Delta B$}
   
   \vspace{0.012\textwidth}
   \hspace{0.79\textwidth}   \color{black} \normalsize {$\it{b}$}

   \vspace{0.025\textwidth}
   \caption{(a) A solar sunspot group's area is filled with mono color. Dotted rectangle shows the area 
used for the calculation. (b) A magnified part of the group in which the individual pixels are seen. 
(c) Positions of the nearest pixels, pixel-a and pixel-b, with respect to the selected pixel-i. 
$\Delta B$ and $\Delta L$ are the latitudinal and longitudinal distances, respectively.}
   \label{F-five}
\end{figure}

Because the pixels in the sunspot group area are black, when such a pixel is encountered in the procession 
of the area, its surrounding eight pixels are taken into account to find the nearest pixels. Both latitudinal 
and longitudinal distances of the eight pixels to the selected pixel are calculated separately. Accepting 
the numerically smallest absolute values as nearest, the nearest two pixels to the selected pixel are found, 
which one is nearest latitudinally (pixel-b) and other is nearest longitudinally (pixel-a). Then, the area 
calculation is performed which was defined graphically in Fig. 5c. $\it{A_P(i)}$, the heliographic area of 
pixel-i is accordingly defined by
\begin{eqnarray} \label{Eq-area}
	       A_P(i) &=& |\Delta B_P(i)| * |\Delta L_P(i)| , \nonumber \\	      
	\Delta B_P(i) &=& P_B(b) - P_B(i) , \nonumber \\ 	      
	\Delta L_P(i) &=& P_L(a) - P_L(i) , \nonumber
\end{eqnarray}
where $\it{P_B(i)}$ and $\it{P_L(i)}$ are the heliographic latitude and longitude of the pixel-i, 
respectively. $\it{P_L(a)}$ is the heliographic longitude of the pixel-a and, $\it{P_B(b)}$ is the heliographic 
latitude of the pixel-b. $\it{\Delta B_P(i)}$ and $\it{\Delta L_P(i)}$ are the latitudinal and longitudinal 
width for the pixel-i, respectively. Since the heliographic area must be a positive value, absolute values 
of the widths have been taken. If the $\it{N}$ is the total number of the pixels forming the sunspot group, 
the total area of the group $\it{A_S}$ will be the sum of the individual pixel areas and is defined by
   \begin{eqnarray} \label{Eq-sum}   	      
	A_S &=& \sum_{i=1}^{N} A_P(i) \nonumber	      
   \end{eqnarray}
\noindent
Since the heliographic coordinates of the pixels were used in all calculations, the effect of the perspective 
on a pixel will be taken into account automatically. Therefore, no correction will be needed for total area 
and the calculated area will be closest to the actual area covered.
\begin{figure}[h]    
   \includegraphics[width=1.0\textwidth,clip=]{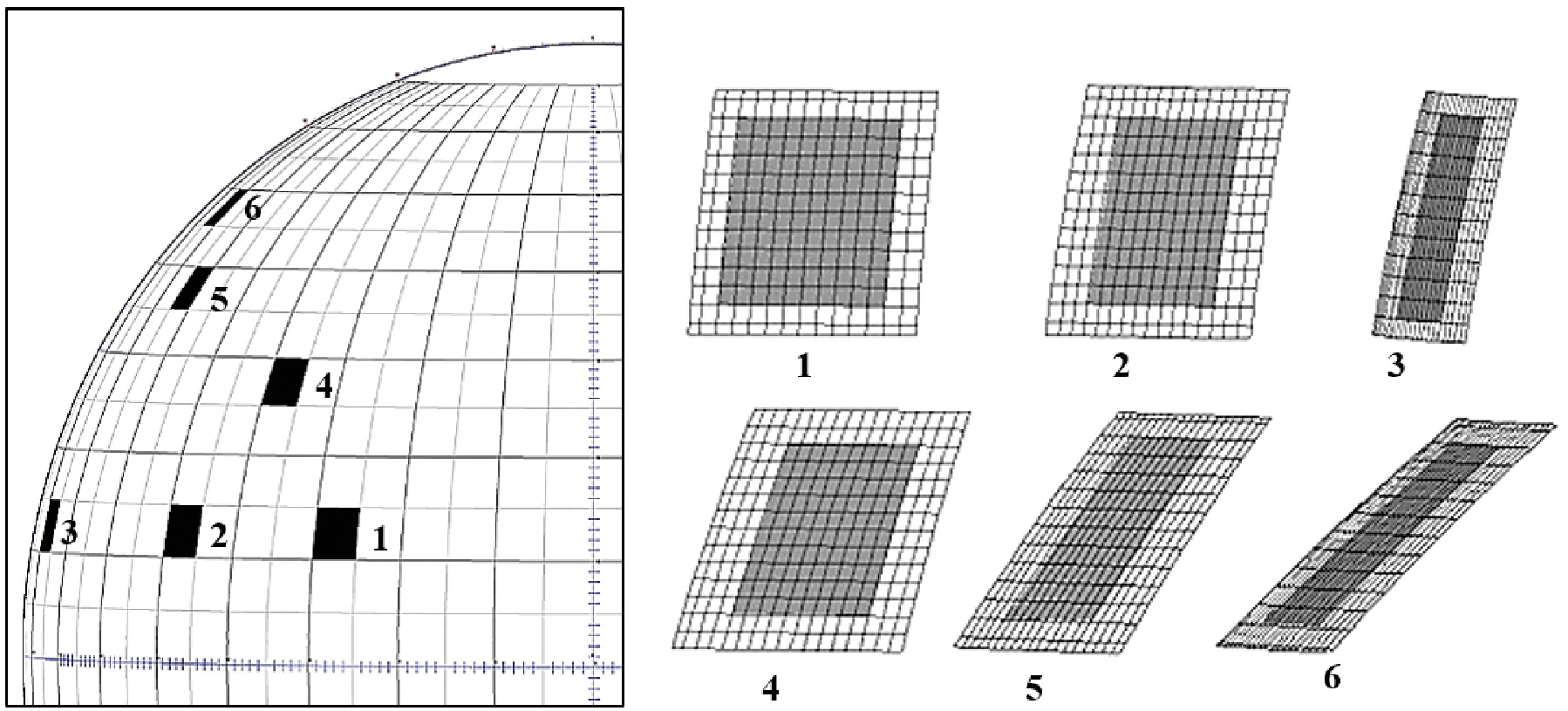}

   \vspace{-0.45 \textwidth}
   \hspace{0.001 \textwidth} \color{black} \small {(a)}
   \hspace{0.38  \textwidth} \color{black} \small {(b)}

   \vspace{0.41\textwidth}

   \caption{(a) Test areas placed at the different latitude and longitude belts. (b) Graticule system 
		drawn at intervals of $0.5^{\circ}$ over test areas.}
   \label{F-six}
\end{figure}
\subsection{Test Measurements} 
  \label{S-test}
A 25$^{\circ}$ squared area which has a width of $5^{\circ}$ in longitude and latitude is marked on an 
empty disk at different longitude and latitude belts as shown in Fig. 6a. All areas are measured individually 
by using the graticule that is formed according to this method (Fig. 6b) and results obtained are listed 
in the Table 1. As seen in the table, the pixel count of the areas close to the edge of the disk are 
relatively small because of the effect of the perspective, but the measured values of the areas are 
very satisfactory, like for the other areas.
\begin{table}[h]
\caption{The measured values of the test areas.}
\centering
\label{T-test}
\begin{tabular}{ccc} 
  \hline
No & Measured value $(^{o})^{2}$ & Number of the pixels in area \\
  \hline
1 & 25.02 & 22,687 \\
2 & 24.96 & 17,298 \\
3 & 24.80 &  5,638 \\
4 & 24.90 & 16,859 \\
5 & 25.08 &  7,924 \\
6 & 25.09 &  3,934 \\
  \hline
\end{tabular}
\end{table}
\begin{figure}[h]     
   \centerline{
     \includegraphics[width=0.998\textwidth,clip=]{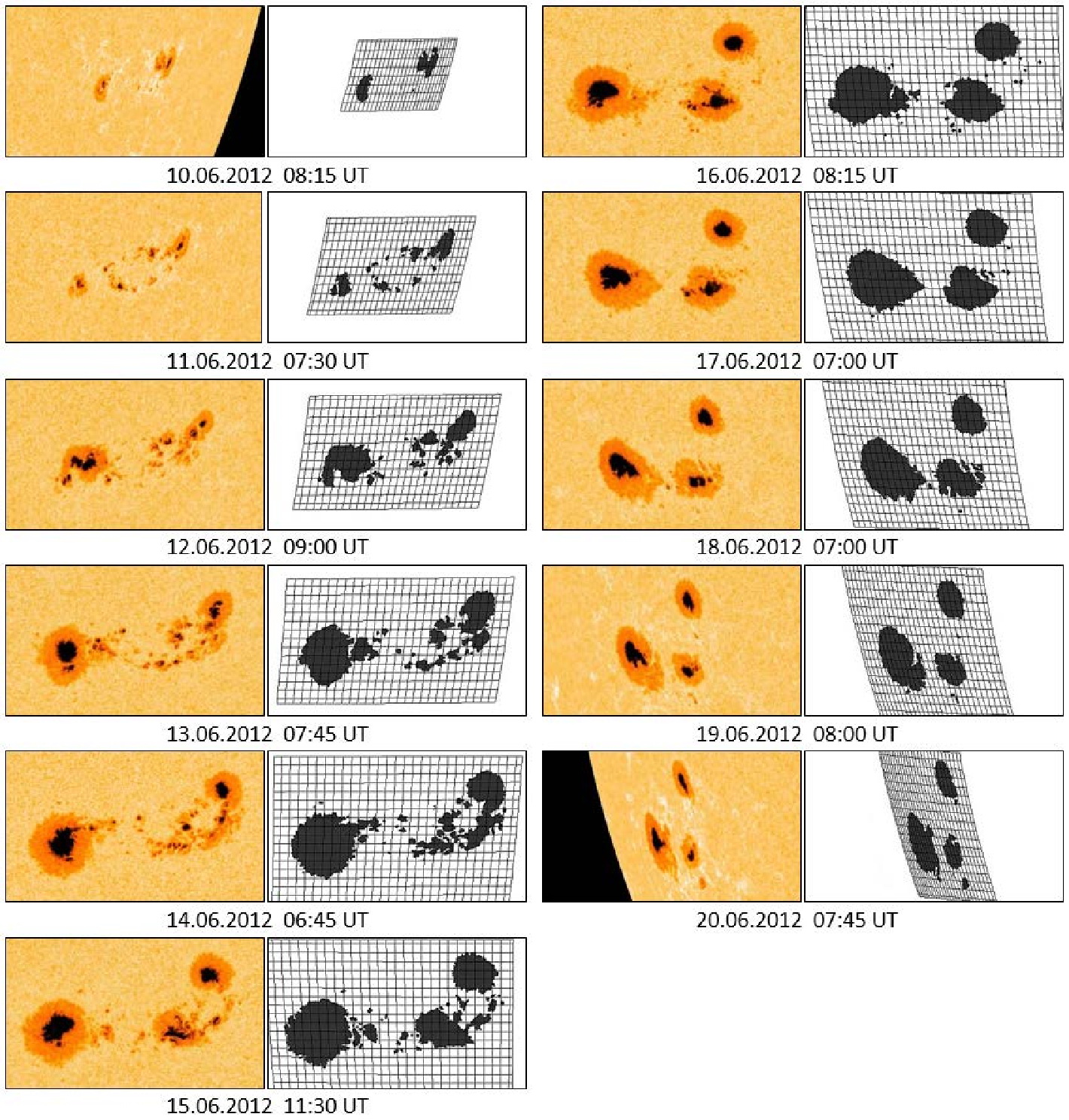}
   }
   \caption{Full disk transition images of the selected sample groups and 
            the positions of the graticule system over the area filled.}
   \label{F-seven}
\end{figure}
\section{Samples of the sunspot area measurements} 
      \label{S-Samples}        
During the selection of a sample in the SDO archive, relatively large and fragmented sunspot groups are 
specially searched. It has been also noted that the sample must have a full transition on the apparent 
solar disk. So the known area variation of the sunspot groups \cite{Gafeira2012} is aimed to be observed 
during the development of the sunspot group. As a result of this, 4096${\times}$4096 pixel sized SDO 
HMIIF solar disk images at 10 - 20.06.2012 have been selected as a sample. 

First, the selected images are converted to the RGB grayscale format, then edge detection of the sunspot 
groups is performed with the Contour Trace process at the threshold intensity level 148, and last manual 
area arrangement and filling procedures are performed. Selected images of the sunspot groups and filled 
areas are shown in Fig. 7. The small part of the graticule is also superimposed over the filled areas to 
show the perspective effect. Measured values of the sample areas are listed in the Table 2 and shown 
graphically in Fig. 8. The change of the observed area is in accordance with the profile of a developing 
sunspot group.

\begin{table}[h]
  \centering
  \caption{The measured values of the sample.}
  \label{T-sample}
  \begin{tabular}{r c c}
     \hline
     Date & Measured Area $(^{o})^{2}$ & Number of the pixels in area\\
     \hline
	2012.06.10 &  5.32 &  2,238 \\
	11 &  6.48 &  3,936 \\
	12 & 12.05 &  9,473 \\
	13 & 18.14 & 16,202 \\
	14 & 22.08 & 21,067 \\
	15 & 24.05 & 23,193 \\
	16 & 22.90 & 21,183 \\
	17 & 22.93 & 19,300 \\
	18 & 22.78 & 16,141 \\
	19 & 21.59 & 11,321 \\
	20 & 21.86 &  7,168 \\
    \hline
  \end{tabular}
\end{table}

\begin{figure}[h]    
   \centering
   \includegraphics[width=0.5\textwidth,clip=]{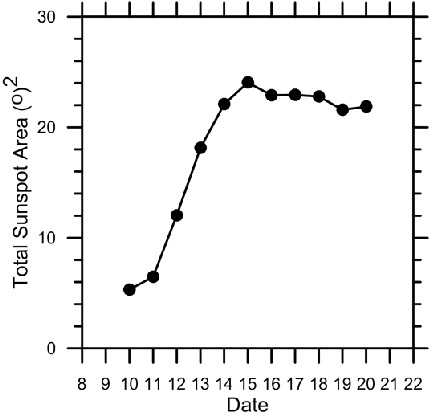}
   \caption{Area change of the sample group with time}
   \label{F-eight}
\end{figure}

\section{Discussion} 
      \label{S-Discussion}        

The principle of the area calculation is based on the pixels of digital images in this method. 
When the size of every pixel is known in heliographic coordinates, the total calculated area 
of the sunspot group will be close to the actual value. Using high-resolution images will increase 
the calculation's accuracy. As a result of this, the pixel area will be small in square degrees 
and the pixel number in the area will increase. Therefore, 4096${\times}$4096 pixel sized, 
high-resolution images of the SDO were used. The diameter of the Sun's disk is 3746 pixels 
in these images. This means that 1 pixel is equal to 1.8 arcminutes in the center of disk and 
5.5 arcminutes around 70$^{\circ}$ longitude on the equator of the Sun. 

Edge detection of the sunspot groups is an important stage of the method. Since the Contour 
Trace algorithm is used, the selection of the threshold intensity value is one of the most 
critical points. Also, since the border of the area is identified visually, the threshold value 
must be selected appropriately. Intensity value of the pixels are in the 0-255 range, experimental 
measurements showed that the amount of the noticeable change can be 5 units above or below the 
proper value. When the threshold value is 5 units below, the areas of sunspot are becoming smaller 
by 2-3\%. This means an area difference of 0.5-0.8 square degrees in a 25$^{\circ}$ squared area, 
because the sunspot group is getting more fragmented and most of the penumbral region is getting 
smaller. On the other hand, when the threshold value is 5 units above, since the borders of the some 
sunspots of the group are merging with each other and penumbral regions are getting wider, the areas 
of sunspot are becoming larger by 4-6\% and this means an area difference of 1-1.5 square degrees 
in a 25$^{\circ}$ squared area. These are the average values and the percentage of the area variation 
strongly depends on the location, size and fragmentation of the sunspot group on the solar disk. 
Especially, if the sunspot group is more fragmented, the change in the percentage will be greater.

The manual re-arrangement of the group area is another critical point. When looking at Fig. 1, some 
difficulties will be seen in which of the bordered areas are belong or not belong the sunspot group. 
However, it is not difficult to decide and this should be done properly, if not, area of the sunspot 
group will be incorrectly calculated. Locating the graticule on the solar disk appropriately is the 
easiest part of the method. But, nevertheless, adjustment should be made carefully. The size of the 
graticule must be equal to the size of the solar disk, and the central point of the graticule must match 
exactly with the central point of the solar disk image. A common error while doing this is to shift 
the graticule a few pixels in any direction, which will be on the order of 2-3 pixels. If the graticule 
is shifted 3 pixels, the sunspot area will change approximately with a 0.1\% in the center and 1-1.5\% 
close to the edge of the disk. When the graticule is shifted to the one direction (for example east 
side), the sunspot groups on that side will be shifted to the center and others on the opposite side 
will be more close to edge of the other side. When a pixel shifts to the center, area of the pixel will 
decrease in square degrees, in the other case, area will increase. Therefore, while the sunspot group 
approaches to the center, the sunspot area will quantitatively decrease, and if the group approaches 
to the edge, the area will increase.

All of the errors mentioned above are the critical points of this method. If the selections 
are made appropriately, the area of the groups will be calculated correctly. More importantly, 
in the other methods, the areas covered by the pixel or small squares are accepted as being equal 
to each other on the solar disk, but they actually are different in size. In the method explained here, 
the areas of the pixels change in square degrees depending on their latitude and longitude. Therefore, 
this is an effective method to calculate areas of the sunspots from digital images of the solar disk. 
Finally, with this method, not only the areas of the sunspots, but also areas of the flares and the 
plages in the chromospheric solar disk images and the areas of the network in Ca II images can be 
calculated. 

\begin{appendix} 
\renewcommand{\thesubsection}{\Alph{subsection}}
\section*{Appendix}
\subsection {Explanations about the area covered by one pixel on the solar image}
The size of the Sun's image changes with the focal length of the telescope used, but it does not change 
on the sky, which is 0.53$^{\circ}$ or 32 arcminutes. In the solar images, only the half of the Sun, 
which is facing us, can be seen. This means that the only 180$^{\circ}$ of the solar sphere is observed.  
When a graticule system is superimposed to the solar disk image, it can be said that one is dealing 
with the surface of the Sun and not with the apparent size of it on the sky. Because the size of the 
graticule depends directly to the size of the solar disk image, any movement on the graticule means 
the movements on the surface of the Sun. From this, it is clearly understood that the distances between 
the latitude lines and longitude lines of the graticule on the surface of the Sun never change, but 
change on the image of the solar disk depending on the perspective.
\begin{figure}[b]    
   \centering
   \includegraphics[width=0.6\textwidth,clip=]{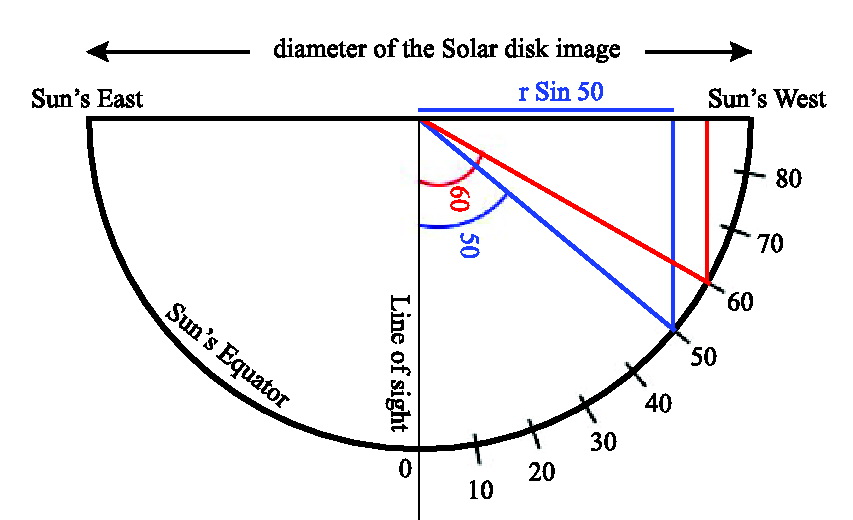}
   \caption{Projection view of the two longitude lines of the graticule at the equator of the Sun, where 
$\it r$ is the radius of the solar disk image. }
   \label{F-nine}
\end{figure}

Some longitude points of the graticule on the equator of the Sun are shown in Fig. 9 for the Western 
hemisphere. As seen in the figure, the projections of the longitude lines depend on the perspective. The 
distances to the disk image center in pixels are proportional to the sine value of the longitude. This 
is given generally by
\begin{eqnarray} \label{Eq-distance}   	      
	D_L &=& \it r \cos B \sin L\nonumber	      
\end{eqnarray}
where $\it D_L$ is the distance of the longitude $\it L$ in pixels to the image center at the latitude 
$\it B$, $\it r$ is the radius of the disk image. The distance between the longitude lines decreases 
while approaching to the edge of the disk, therefore the number of the pixels also decreases, but increases 
towards the center. For example, at the equator ($\it B$ = 0), the number of the pixels ($\it NP$) for a 
width of 1$^{\circ}$ in the center and near to the edge of the image will be given by
\begin{eqnarray} \label{Eq-dist-sample}  	      
	 D_1 &=& {\it r} \sin 1^{\circ}, \hspace{0.31in}  
         D_{2} = {\it r} \sin 2^{\circ}, \hspace{0.2in} NP_{center} = D_{2} - D_1\nonumber \\
	 D_{70} &=& {\it r} \sin 70^{\circ}, \hspace{0.2in}  
         D_{71} = {\it r} \sin 71^{\circ}, \hspace{0.24in} NP_{edge} = D_{71} - D_{70}\nonumber \\
         NP_{center} &=& {\it r} (\sin 2^{\circ} \hspace{0.06in} -  
	\sin 1^{\circ}\hspace{0.06in} ) = 0.0174 {\it r} \nonumber \\         
	NP_{edge} &=& {\it r} (\sin 71^{\circ} - \sin 70^{\circ}) = 0.0058 {\it r} \nonumber	      
\end{eqnarray}
The SDO images used in this work has a radius of 1873 pixels and the number of the pixels in these images 
in 1$^{\circ}$ width are $NP_{center}$ = 33px and $NP_{edge}$ = 11px. One pixel is equal to 1.8 
arcminutes (60$'$/33px) in the center and 5.5 arcminutes (60$'$/11px) near to the edge of disk. 

Similarly, the size of a pixel in kilometers can be calculated by using the perimeter of the Sun in 
kilometers. Since the diameter of the Sun is 1 396 400 km, the perimeter of the Sun will be 4 386 920 km. 
Hence, 1$^{\circ}$ will correspond to 12 186 km and the 1 arcminute will correspond to 203 km. As seen 
from this calculation, the area covered by the pixels on the Sun's surface can be calculated in square-kilometers 
as well as the area of the sunspot groups. 

\subsection {Comprehensive informations about the finding the nearest pixels}
As explained in Section 2.1, every pixel forming the sunspot group have heliographic coordinates according 
to the graticule system superimposed on it. While processing the area of the group, each pixel of the group 
individually checked to find the nearest two pixels to it. To do this, the surrounding the pixels of the 
selected pixel are taken into account. A shematic view of this situation is shown in Fig. 10. Although the 
lines of the latitude and longitude of the graticule are curved, since they are on the surface of the sphere, 
these are shown as straight lines for simplicity. 

Let the selected pixel (No. 0) be at the heliographic coordinates $\it B_0$ and $\it L_0$ as shown in 
Fig. 10. Two pixels (4 and 5) are the same latitude and other two (2 and 8) are the same longitude with 
the selected pixel. But this is not correct for all situations. Specifically, when the sunspot group is 
close to the edge of the disk, the longitudinal lines of the graticule become more inclined and the corner 
pixels shift in longitude and become the same longitude with the selected pixel rather than the upper and 
lower pixels. Pixels (3, 5 and 9) on the right of the selected pixel have smaller, pixels on the left (1, 4 
and 7) have larger longitudes than the selected pixel's. Also pixels (7, 8 and 9) below the selected pixel 
have smaller, upper of them (1, 2 and 3) have larger latitudes than the selected pixel's.    
\begin{figure}[h]    
   \centering
   \includegraphics[width=0.5\textwidth,clip=]{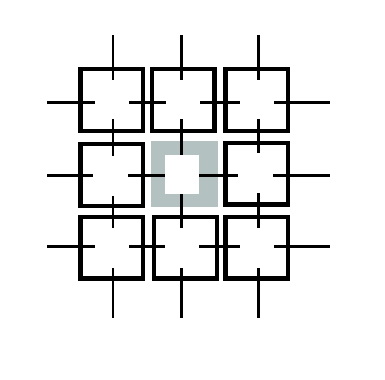}

   \vspace{-0.387\textwidth}
   \hspace{0.12 \textwidth} \color{black} \normalsize {1}
   \hspace{0.06 \textwidth} \color{black} \normalsize {2}
   \hspace{0.06 \textwidth} \color{black} \normalsize {3}
   \hspace{0.08 \textwidth} \color{black} \normalsize {B+}

   \vspace{0.057\textwidth}
   \hspace{0.11 \textwidth} \color{black} \normalsize {4}
   \hspace{0.06 \textwidth} \color{black} \normalsize {0}
   \hspace{0.06 \textwidth} \color{black} \normalsize {5}
   \hspace{0.08 \textwidth} \color{black} \normalsize {$B_0$}

   \vspace{0.058\textwidth}
   \hspace{0.11 \textwidth} \color{black} \normalsize {7}
   \hspace{0.06 \textwidth} \color{black} \normalsize {8}
   \hspace{0.06 \textwidth} \color{black} \normalsize {9}
   \hspace{0.09 \textwidth} \color{black} \normalsize {B-}

   \vspace{0.08\textwidth}
   \hspace{-0.02 \textwidth} \color{black} \normalsize {L+}
   \hspace{0.04 \textwidth} \color{black} \normalsize {$L_0$}
   \hspace{0.04 \textwidth} \color{black} \normalsize {L-}

   \caption{The surrounding pixels of the selected pixel (No. 0). $\it {B+}$ and $\it {L+}$ represents 
larger values, and $\it {B-}$ and $\it {L-}$ represents smaller values of the latitude and the longitude 
with respect to $\it {B_0}$ and $\it {L_0}$.}
   \label{F-ten}
\end{figure}

In order to find the nearest pixels, both the latitudinal and longitudinal differences in degrees of the 
surrounding the pixels to the selected pixel must be calculated separately. These differences are given 
by
\begin{eqnarray} \label{Eq-diffrence}  	      
	 {\Delta B}_i &=& B_0 - B_i, \hspace{0.31in} B_i < B_0 \Rightarrow {\Delta B}_i > 0, 
	\hspace{0.31in} B_i > B_0 \Rightarrow {\Delta B}_i < 0 , \nonumber \\ 
	 {\Delta L}_i &=& L_0 - L_i, \hspace{0.34in} L_i < L_0 \Rightarrow {\Delta L}_i > 0, 
	\hspace{0.34in} L_i > L_0 \Rightarrow {\Delta L}_i < 0 , \nonumber
\end{eqnarray}
where $\it i$ is the position number of the surrounding pixels and has values from 1 to 8, respectively. 
${\Delta B}_i$ is the latitudinal difference, and ${\Delta L}_i$ is the longitudinal difference for 
the number $\it i$ pixel, $B_i$ is the latitude of the number $\it i$ pixel, $\it L_i$ is the longitude 
of the $\it i$th pixel. As seen from the equations, $\Delta B$ and $\Delta L$ may accordingly have 
negative or positive values. Since we are not dealing with their signs, the absolute values or 
numerical values of them must be taken into account to find the smallest for both latitudinal and 
longitudinal differences. After the smallest $\Delta B$ and the smallest $\Delta L$ are found, the 
area of the selected pixel is calculated as the product of these two values.

\end{appendix} 

\begin{acknowledgements}
Author thanks to his colleague Res. Asst. Dr. Asuman G\"{u}ltekin from the Istanbul University Science 
Faculty, Astronomy and Space Sciences Department for his contributions towards developing the method. 
Also, thanks to Res. Asst. Ba\c{s}ar Co\c{s}kuno\u{g}lu for his contributions towards improving the 
language of the manuscript. Also, thanks to anonymous reviewer for his/her valuable suggestions and 
comments improving manuscript. This work was supported by Scientific Research Projects Coordination 
Unit of Istanbul University with the project number 24242 and the project number 6021.
\end{acknowledgements}



\end{document}